\documentclass[letterpaper]{article}
\usepackage{aaai2026}
\usepackage{times}
\usepackage{helvet}
\usepackage{courier}
\usepackage[hyphens]{url}
\usepackage{graphicx}
\usepackage{natbib}
\usepackage{caption}
\usepackage[most]{tcolorbox}
\usepackage{placeins}
\usepackage{float}

\usepackage{xcolor}

\title{The 2026 World Cup, Anime Edition: Production, Engagement, \\ and Circulation of Generative Memetic Sports Content}

\author{
    Azneef Chowdhury,
    Ho-Chun Herbert Chang
}

\affiliations{
    Program in Quantitative Social Science, Dartmouth College\\
    Hanover, NH 03755\\[1em]
    azneef.a.chowdhury.29@dartmouth.edu,
    herbert.chang@dartmouth.edu
}

\begin{document}
\nocopyright
\maketitle

\begin{abstract}
Generative AI gives audiences new ways to portray real sporting figures through familiar fictional worlds. Expanded opportunities to create, however, need not broaden cultural representation or distribute attention evenly. We examine this tension through the supply, demand, and circulation of anime-style football videos on Instagram. Using 2,021 posts and associated video and account data, we combine caption entity extraction, human-reviewed visual identity clustering, engagement analysis, and content-reuse networks. Representation is uneven across athletes, while \emph{Blue Lock}, \emph{Naruto}, and \emph{One Piece} account for 60.1\% of explicit franchise mentions. Among posts referencing generative tools, rankings by total engagement differ from rankings by average engagement per video, distinguishing aggregate visibility from response to individual posts. Reuse is extensive: 772 of 1,548 evaluated videos belong to 124 detected groups, including 71 spanning multiple known accounts. Shared material connects a large account component alongside smaller groups, and later observed uploads receive lower median normalized engagement than first observed appearances. These findings characterize generative cultural crossover as a practice organized around concentrated cultural resources, uneven attention, and extensive recirculation. More broadly, they distinguish opportunities for visual authorship from the processes through which particular representations become widely visible.
\end{abstract}

\section{Introduction}

The FIFA World Cup brings global audiences together around a shared sequence of matches, rivalries, and sporting figures \citep{fifa2024audience}. These events also provide material for cultural production beyond the broadcast, where fans reinterpret athletes and their performances through familiar fictional worlds. Generative AI extends this practice by enabling newly synthesized scenes rather than relying solely on existing footage.

This practice builds on established traditions of audiovisual remix and sports fan fiction \citep{ito2010rewards,waysdorf2015football}. Generative AI adds synthetic image and video production to this repertoire, extending the relationship between cultural participation and human curation. Generative memesis describes how customized AI visuals become part of meme production and communication \citep{chang2026meme}. During Taiwan's 2024 Bluebird Movement, AI-generated imagery likewise served as a resource for collective expression and mobilization \citep{weener2026bluebird}. Football--anime crossover extends these concerns to a cultural practice organized around sporting celebrities and shared fictional worlds.

Expanded opportunities to create, however, do not necessarily produce broader representation or more evenly distributed attention. Many generated depictions can center on the same athletes and franchises, while repeated uploads can increase visibility without adding distinct material. These possibilities intersect with platform-mediated cultural production, in which content is selected, repackaged, and circulated under unequal conditions of visibility \citep{nieborg2018platformization}. The central question is therefore not simply what generative tools enable people to make, but how the resulting cultural supply relates to audience attention and cross-account circulation.

We examine 2,021 Instagram posts and their associated videos collected through Meta's Content Library. The analysis connects three dimensions: \emph{supply}, characterizing represented athletes, anime references, and named production tools; \emph{demand}, examining recorded engagement; and \emph{circulation}, mapping shared material across accounts and time. We combine caption analysis, human-reviewed visual identity labels, engagement measures, and reuse detection. The sampled content centers on leading football figures, while three anime franchises account for 60.1\% of franchise mentions. Detected reuse groups contain 772 of 1,548 evaluated videos, and later observed appearances receive lower median normalized engagement. Together, these findings connect generative cultural production to concentrated representation, selective attention, and extensive recirculation. The contribution is a linked account of how real-person depictions become visible cultural objects, rather than an analysis of generation or engagement alone.

\section{Related Work}

\subsection{Generative Cultural Production}

Fan production transforms commercial media through practices organized around shared conventions, cultural knowledge, and recognition. Anime music video communities combine open participation with distinctions in reputation and expertise, demonstrating that access to creative tools does not eliminate hierarchies among contributors \citep{ito2010rewards}. Generative AI extends these practices through relationships among creators, communities, and technical systems, rather than isolated interactions between individual users and software \citep{lipang2024fandom}. Generative memesis further distinguishes the production of customized synthetic imagery from the human selection and curation through which that imagery becomes communication \citep{chang2026meme}.

Generative cultural production also encompasses practices beyond noncommercial fandom. Ethnographic research on AI-video producers in China between 2023 and 2025 documents workflows organized around testing formats, adapting successful templates, and coordinating multiple accounts, with speed and scale prioritized over a stable authorial identity \citep{tang2026slop}. These practices place creative expression alongside strategic efforts to secure attention. The relevant object of study is therefore a production ecosystem, not simply a collection of independently authored works. In football--anime content, represented figures, fictional references, and named tools characterize the visible resources of this ecosystem, while engagement and reuse reveal different aspects of its organization.

\subsection{Sport, Anime, and Cultural Crossover}
Sporting performances and athletes' public lives already provide material for transformative storytelling. Football fan fiction treats matches, team relationships, and celebrity publicity as a source text that can be selectively developed and reinterpreted \citep{waysdorf2015football}. Real-person fiction distinguishes the public figure, their mediated persona, and the character collectively constructed by fans \citep{popova2017persona}. This distinction matters for synthetic portrayals: identifying the athlete answers who is represented, but not what fictional role or interpretation accompanies that representation. Such portrayals raise questions about expressive agency and control over public characterization even when they are not presented as documentary evidence.

Anime supplies a repertoire of characters and conventions that can travel beyond their original stories. Its media mix connects character circulation across animation, other media, and commodities \citep{steinberg2012mediamix}; recurring visual tropes and personality archetypes also structure character design and are associated with online popularity \citep{liu2026anime}. Exchanges between sport and anime predate generative AI, including adaptations of figure-skating choreography between real performances and \emph{Yuri!!! on Ice} \citep{sottile2025fanception}. Conceptual blending offers a framework for understanding how elements from different domains combine into new interpretations \citep{fauconnier1998integration}. In football--anime crossover, sporting reality supplies identities and events, while anime supplies fictional resources for their reinterpretation. This perspective motivates examining which figures and franchises dominate the sampled content, while distinguishing the use of an anime aesthetic from explicit character references and event-to-storyline mappings.

\subsection{Cultural Supply, Audience Demand, and Unequal Attention}

The availability of cultural material and its distribution of attention are distinct outcomes. Social influence increased the inequality and unpredictability of success in an experimental music market, demonstrating that popularity need not directly reflect independent evaluations of quality \citep{salganik2006inequality}. Generative assistance introduces a related tension: in a short-story experiment, access to AI-generated ideas improved individual evaluations while increasing similarity across the resulting stories \citep{doshi2024creativity}. These findings motivate separating creative opportunity from collective diversity. Platformization adds another layer, making cultural production responsive to audience metrics, distribution infrastructures, and changing platform rules \citep{nieborg2018platformization}. A larger supply of depictions can therefore coexist with concentrated representation and unequal visibility.

Recorded engagement also depends on the relationship between content and its presentation. Multimodal popularity research distinguishes textual and visual features from contextual factors such as timing and creator identity \citep{hessel2017cats}, while text and images can contribute complementary rather than redundant information \citep{vempala2019relationship}. Synthetic provenance alone is likewise insufficient to explain attention. In Instagram imagery surrounding the 2024 U.S. presidential election, meme format predicted engagement more strongly than AI generation alone, with an additional positive association for their interaction \citep{chang2026meme}. During Taiwan's 2024 Bluebird Movement on Threads, human photographs and AI-generated symbols occupied different communicative roles, and exposure did not consistently align with active engagement \citep{weener2026bluebird}. These findings motivate examining how sporting identities, cultural references, and posting context relate to attention. We use \emph{demand} to describe recorded viewing and engagement, rather than treating these measures as direct evidence of preference or approval.

\subsection{Circulation, Imitation, and Originality}

Cultural participation includes selecting and redistributing material, not only producing it. Spreadability emphasizes the value and meaning generated through audience decisions to circulate media \citep{jenkins2013spreadable}. Internet memes likewise develop through the reproduction and transformation of content, form, and communicative stance \citep{shifman2013memes}. Platform features can organize participation around imitation, as conceptualized by \emph{imitation publics} on TikTok \citep{zulli2022imitation}. These perspectives make repetition a substantive feature of cultural life rather than merely a data-cleaning problem. Reusing a fictional convention to create a new depiction nevertheless differs from reposting substantially the same material.

Circulation also creates unequal opportunities for attention. Resubmitted images provide a way to examine associations with titles, communities, timing, and previous appearances while partially holding the underlying content constant \citep{lakkaraju2013name}. Platforms impose their own distinctions: in 2024, Instagram announced that it would favor the original when recommending content it identified as identical \citep{instagram2024originality}. This stated policy places originality alongside practices in which redistribution helps make cultural material shared, without establishing how any particular post was ranked. Linking content reuse to identities, accounts, and engagement allows us to distinguish breadth of representation from repeated visibility. It also separates observed patterns of shared material from claims about original authorship or direct copying. Together, these distinctions connect the supply of generative depictions to the networks through which they recur and the attention their successive appearances receive.

\subsection{Research Questions}

With an ecology of attention surrounding these dimensions, we formulate the following research questions:

\textbf{RQ1 (Supply):} Which sporting figures, anime references, and named generative tools characterize anime-style football videos, and how concentrated is this repertoire?

\textbf{RQ2 (Demand):} Which content and posting characteristics are associated with recorded engagement?

\textbf{RQ3 (Circulation):} How does reused material recur across accounts and over time, and how does engagement differ across successive appearances?

\section{Methods}

\subsection{Data and Corpus}

\begin{figure}[ht]
    \centering
    \includegraphics[width=\linewidth]{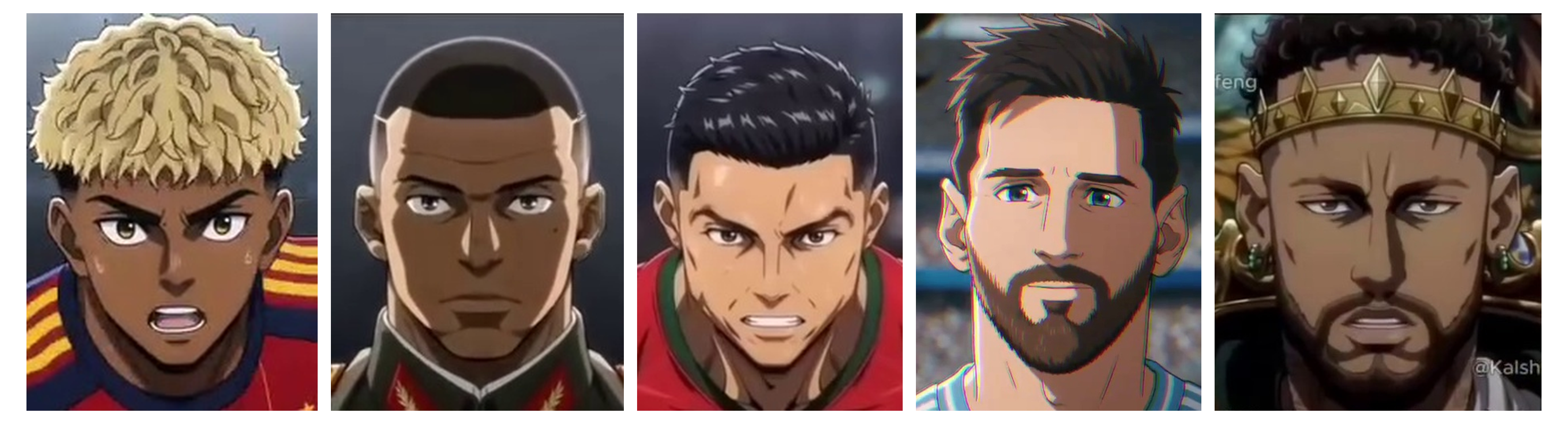}
    \caption{Examples of visually identified football figures in the corpus. From left to right: Lamine Yamal, Kylian Mbappé, Cristiano Ronaldo, Lionel Messi, and Neymar Jr.}
    \label{fig:players}
\end{figure}

We collected posts through Meta's Content Library
\cite{meta2023contentlibrary} between April 1 and August 15,
2026, using the search terms \textit{anime} and
\textit{world cup}. This period spans the lead-up to, duration of, and immediate aftermath of the 2026 FIFA World Cup. The source dataset contains 2,021 unique posts with captions, hashtags, timestamps, account information, and engagement measures including views, likes, and comments. We downloaded 1,901 associated video files for visual analysis.
To assess whether the search retrieved the generative content targeted by the study, we manually reviewed a random sample of $100$ videos. All sampled videos were AI-generated.

We separately collected account-level metadata through
Social Blade \cite{socialblade2026}, including follower count,
following count, and total posts. Out of 833 attempted account lookups, 811 returned successfully. Follower counts ranged from 29 to approximately 39.1 million, with a median of approximately 66,700.

Different analyses use different subsets of the corpus according to data availability. Visual identity analyses require usable face detections and reviewed identity clusters; reuse analyses include 1,548 videos with the necessary face crops; and engagement analyses require matched post- and account-level metadata. We report the relevant sample size with each analysis.

\subsection{Visual Identity Measurement}

We developed a human-in-the-loop pipeline to identify recurring
football figures across stylized videos. YOLO11
\cite{jocher2024yolo11} and BoT-SORT
\cite{aharon2022botsort} were first used to detect and track
people within each video. Tracks appearing for fewer than five
frames were discarded, and up to 15 approximately evenly spaced
frames were sampled from each remaining track.

Faces were detected within person crops using SCRFD
\cite{guo2022scrfd}. We retained detections meeting quality
criteria based on face size, image sharpness, and estimated
head pose. To reduce redundancy, candidate faces were ranked using person-detection confidence, face-detection confidence, and face area,

\begin{equation}
S_i =
C_{\mathrm{person},i}
C_{\mathrm{face},i}
\sqrt{A_i},
\end{equation}

and filtered for temporal and perceptual diversity. At most five faces were retained per track.

Selected faces were aligned using five facial landmarks and
embedded with ArcFace \cite{deng2019arcface} into
L2-normalized 512-dimensional vectors. We clustered embeddings
across videos in two stages. HDBSCAN
\cite{mcinnes2017hdbscan} first identified locally coherent
micro-clusters using Euclidean distance, a minimum cluster size
of three, and \texttt{min\_samples=1}. We then represented each
micro-cluster using the normalized centroid of its most typical
50\% of members and merged these representatives using
complete-link hierarchical clustering with cosine distance.

The retained solution contained 900 visual clusters, of which we manually labeled 500. These 500 reviewed clusters covered 1,370 unique videos, corresponding to 88.5\% of the 1,548 videos with clustered face crops and 72.1\% of all 1,901 videos. We stopped manual labeling at 500 clusters because the remaining clusters provided limited marginal video coverage: the 400 unreviewed clusters contained a median of four face tracks but only one unique video, and although they collectively touched 465 videos, only 103 of those videos were not already represented by a reviewed cluster. Thus, additional cluster labeling would have substantially increased annotation effort while adding relatively little coverage of previously unseen videos.

Observations initially labeled as HDBSCAN noise were reassigned when their cosine similarity to a final cluster representative was at least 0.45. 500 of the 900 resulting clusters were manually reviewed and labeled when a football figure could be identified. Multiple clusters assigned to the same figure were combined in downstream player-level analyses. Figures without a reviewed label are treated as not identified rather than visually absent.

\paragraph{Human Validation of Player Labels.}
To evaluate the reliability of the manually assigned player identities, we conducted a separate human-validation step. For each of the 18 identified players, we randomly sampled up to 30 face images from the corresponding labeled clusters and manually verified whether each image depicted the assigned player. Across 524 sampled faces, 502 were correctly labeled, yielding an overall accuracy of 95.8\% (95\% CI: 93.7--97.2\%). These results indicate that the labeled face clusters generally provide high-precision player identification.

\subsection{Caption and Cultural-Reference Extraction}

We separately analyzed the original Instagram captions and
hashtags using GPT-OSS 120B \cite{openai2025gptoss}. Each post
was processed independently at temperature 0 using a structured
prompt that instructed the model to extract only explicitly
mentioned entities and to normalize common aliases such as
``Messi'' to ``Lionel Messi'' and ``PSG'' to
``Paris Saint-Germain.''

The extraction returned 13 fields: AI model names, model versions, player names, countries, anime titles, anime characters, animation studios, club teams, match events, tournament rounds, source handles, managers or pundits, and venues. The prompt explicitly distinguished real football players from anime characters, national teams from club teams, and managers or pundits from players. Tool and franchise mentions are treated as textual references and do not independently establish which model generated a video or which franchise is visually represented.

\subsection{Engagement Analysis}

We use recorded views, likes, and comments as engagement outcomes. Because accounts differ substantially in audience size and posts differ in age, multivariate models include $\log(1+\mathrm{followers})$ as controls.

For each outcome $Y_i$, we estimate ordinary least squares models of $\log(1+Y_i)$. We compare specifications containing caption-category indicators, visually identified football figures, and both sets of measures together. Visual identity indicators are included only when sufficient positive and negative observations are available for estimation. Standard errors are clustered by account. These models estimate conditional associations rather than causal effects.

We also compare aggregate and mean engagement among posts explicitly mentioning different generative tools. These descriptive comparisons characterize posts associated with named tools rather than the performance of the underlying systems themselves.

\subsection{Content Reuse and Sharing Networks}

To identify repeated visual material, we used the face crops produced by the visual pipeline. The reuse analysis contains 19,094 selected face crops from 1,548 videos. Each crop was represented using a 64-bit perceptual hash (pHash). For each video pair, we calculated bidirectional best-match Hamming distances between crops and classified videos as linked when their minimum bidirectional match coverage at Hamming distance four was greater than zero. Connected components containing at least two videos define detected reuse groups.

These groups identify shared visual material rather than verified whole-video duplicates, original authorship, or direct copying. Because connected components are transitive, two videos can also belong to the same group through an intermediate match.

We use the detected groups to examine circulation across accounts and time. For each group, accounts are ordered using their earliest observed posting timestamp. Directed account-level edges connect earlier-posting accounts to later-posting accounts within the same group, and edge weights count the number of distinct groups supporting each relationship. These edges represent temporal co-sharing rather than observed transmission paths.

To characterize circulation speed, we calculate the interval between the earliest and latest known timestamps within each group. We also compare engagement across successive appearances of the same detected material. Posts are ordered as first, second, and third-or-later observed appearances within each eligible group. Engagement is normalized for account audience size and expressed relative to the first observed appearance. ``First'' therefore refers to the earliest post observed in our corpus and does not establish original authorship.

\section{Results}

The posts in our corpus accumulated approximately 490 million recorded views. We organize the results around three dimensions of the cultural ecosystem: the supply of representations and cultural references, the demand reflected in recorded engagement, and the circulation of shared material across accounts.

\subsection{Supply: Cultural Production}

\begin{figure}[ht]
    \centering
    \includegraphics[width=\linewidth]{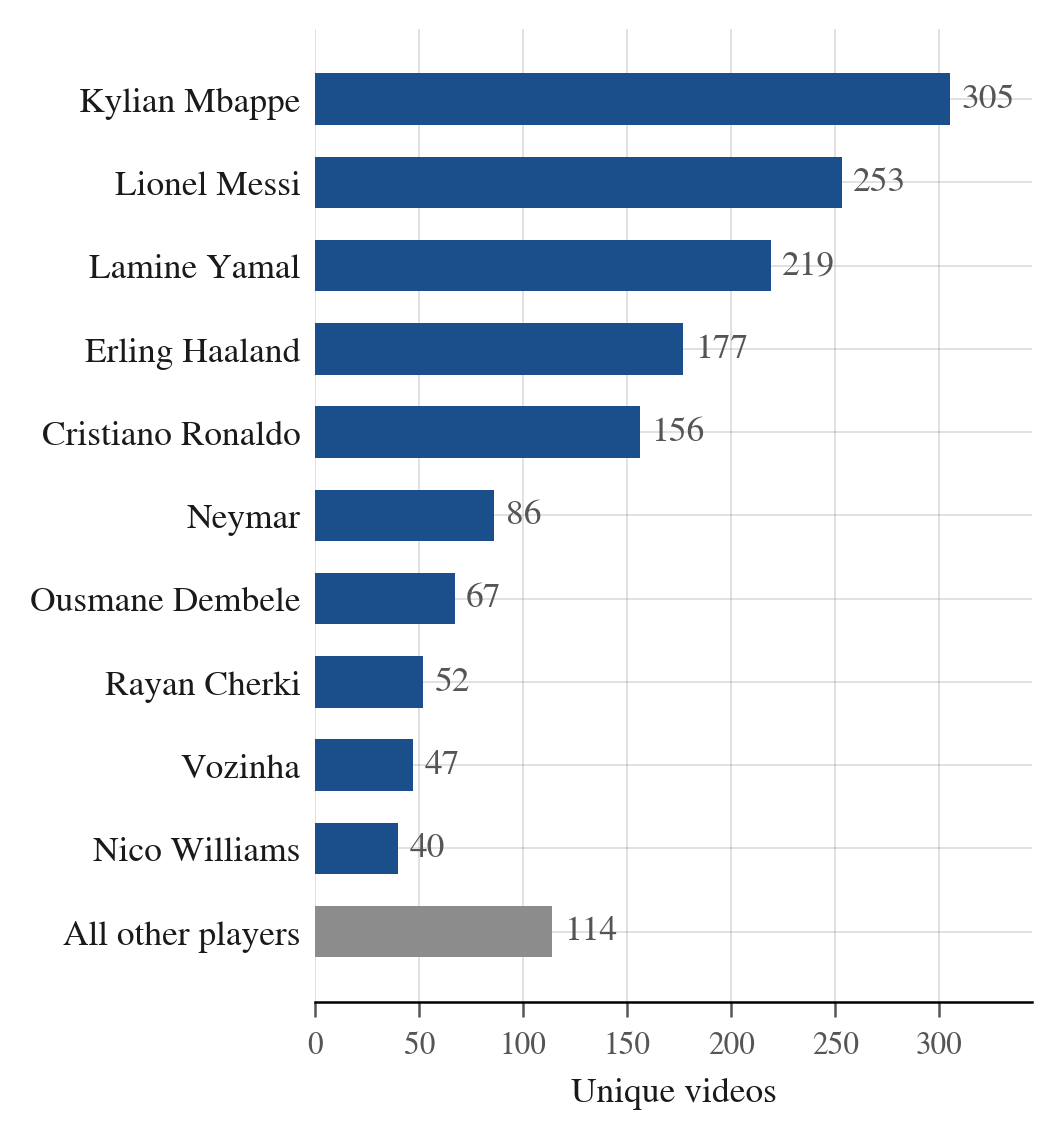}
    \caption{Visual representation of football figures in the sampled videos. Bars report the number of unique videos containing each identified player. The ten most frequently represented players are shown separately, with remaining identified players aggregated into ``All other players.''}
    \label{fig:player_coverage}
\end{figure}

We first examined how frequently individual players appeared in the recovered visual identity clusters. Among the 18 players manually reviewed and assigned to identity clusters, coverage varied substantially across players. Kylian Mbappé appeared across 305 unique videos, followed by Lionel Messi (253), Lamine Yamal (219), Erling Haaland (177), and Cristiano Ronaldo (156). Coverage declined substantially among less frequently represented players, with Bruno Fernandes appearing in only three videos and Vitinha in seven (Figure~\ref{fig:player_coverage}).

This distribution shows that visual representation is concentrated around a relatively small set of prominent football figures rather than distributed evenly across identified players. At the same time, post-level prevalence can reflect both the production of distinct depictions and the repeated circulation of the same material. We return to this distinction in the circulation analysis below.

\begin{figure}[H]
    \centering
    \includegraphics[width=\linewidth]{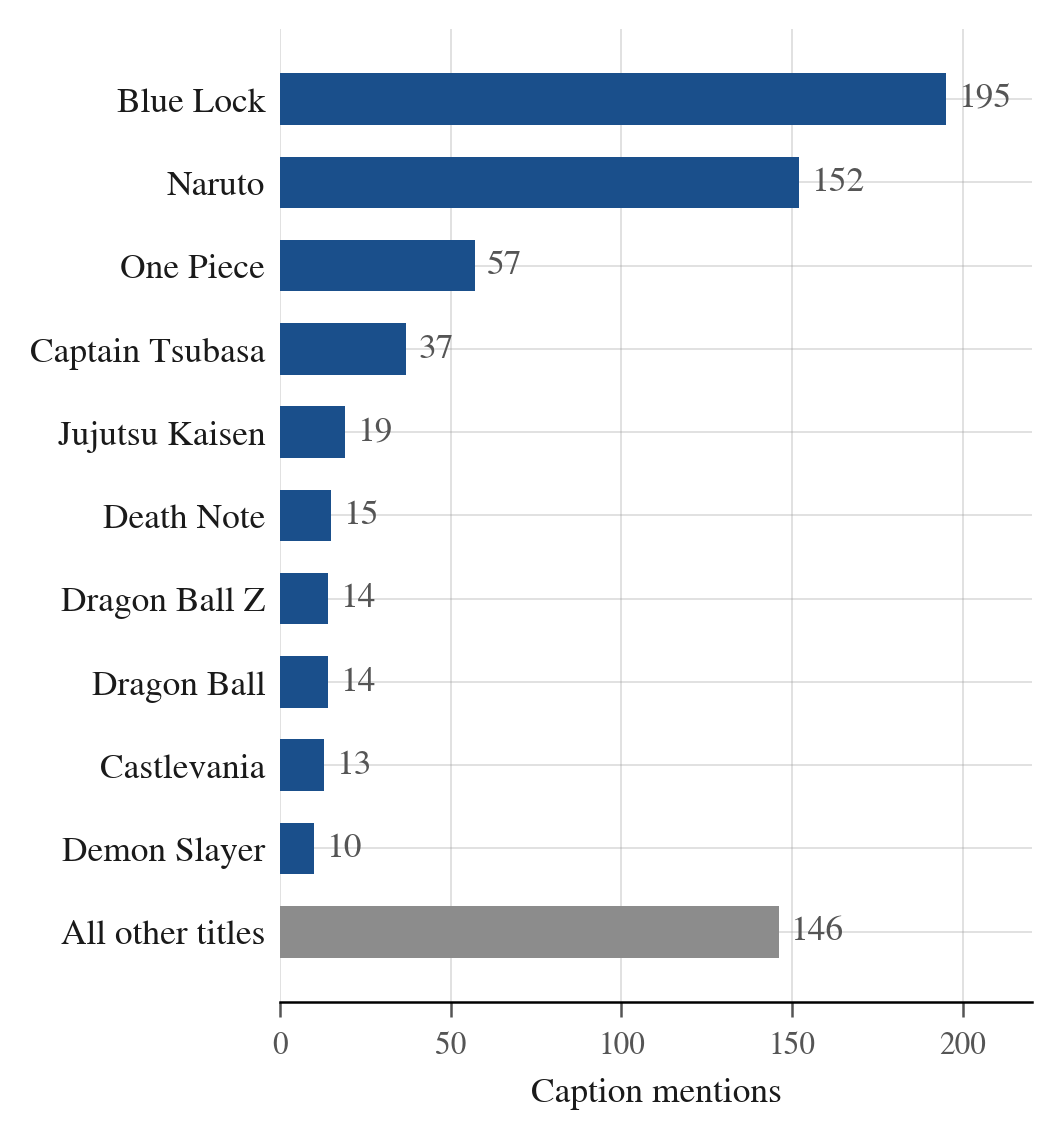}
    \caption{Anime franchise mentions extracted from post captions and hashtags ($n=672$ mentions). Titles outside the ten most frequent are aggregated into ``All other titles.''}
    \label{fig:franchises}
\end{figure}

We also examined which anime franchises were explicitly referenced in post captions. Across 672 franchise mentions, \textit{Blue Lock} was the most frequently referenced title, appearing 195 times (29.0\%), followed by \textit{Naruto} with 152 mentions (22.6\%) and \textit{One Piece} with 57 (8.5\%). Other frequently referenced franchises included \textit{Captain Tsubasa} (37), \textit{Jujutsu Kaisen} (19), \textit{Death Note} (15), and \textit{Demon Slayer} (10). The remaining 146 mentions (21.7\%) were distributed across a long tail of less frequently named titles (Figure~\ref{fig:franchises}).

\begin{figure*}[!t]
    \centering
    \includegraphics[width=0.82\textwidth]
    {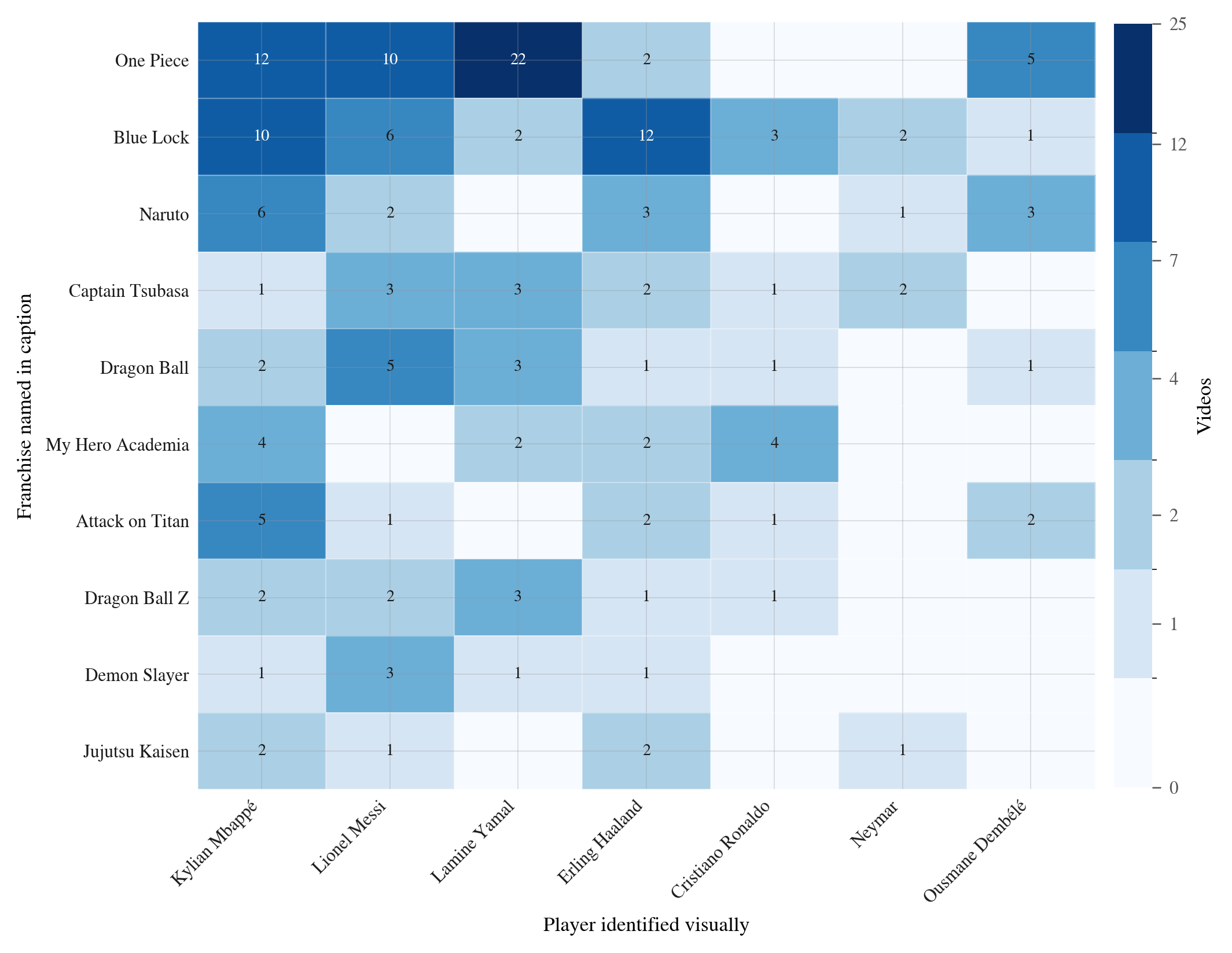}
    \caption{Video-level co-occurrence between visually identified
    football players and explicitly mentioned anime franchises.
    Cells report the number of videos containing both the identified
    player and the caption-derived franchise mention. Co-occurrence
    does not independently establish that the player was rendered in
    that franchise's style or fictional setting.}
    \label{fig:player_franchise_overlap}
\end{figure*}

We next examined how caption-mentioned anime franchises overlapped with visually identified football players.  Figure~\ref{fig:player_franchise_overlap} cross-tabulates the two measures at the video level. The most frequent pairing was Lamine Yamal with \textit{One Piece}, appearing in 22 videos. \textit{One Piece} also frequently co-occurred with Kylian Mbappé (12 videos) and Lionel Messi (10), while \textit{Blue Lock} most frequently appeared alongside Erling Haaland (12) and Mbappé (10). Less frequently mentioned franchises were distributed more sparsely across players. These results suggest that the football--anime crossover is organized around recurring player--franchise combinations rather than an even mixing of sporting and fictional identities.

The distribution is heavily concentrated: the three most common titles account for 60.1\% of all franchise mentions. The prominence of \textit{Blue Lock} is particularly notable given its direct association with football, but the prominence of \textit{Naruto} and \textit{One Piece} shows that the crossover also draws heavily on fictional worlds unrelated to football. Generative cultural production in this corpus therefore combines a broad range of possible tools and references with a comparatively narrow set of highly visible athletes and franchises. This answers \textbf{RQ1}.

\subsection{Demand: Engagement}

\begin{figure*}[t]
    \centering
    \includegraphics[width=1\linewidth]{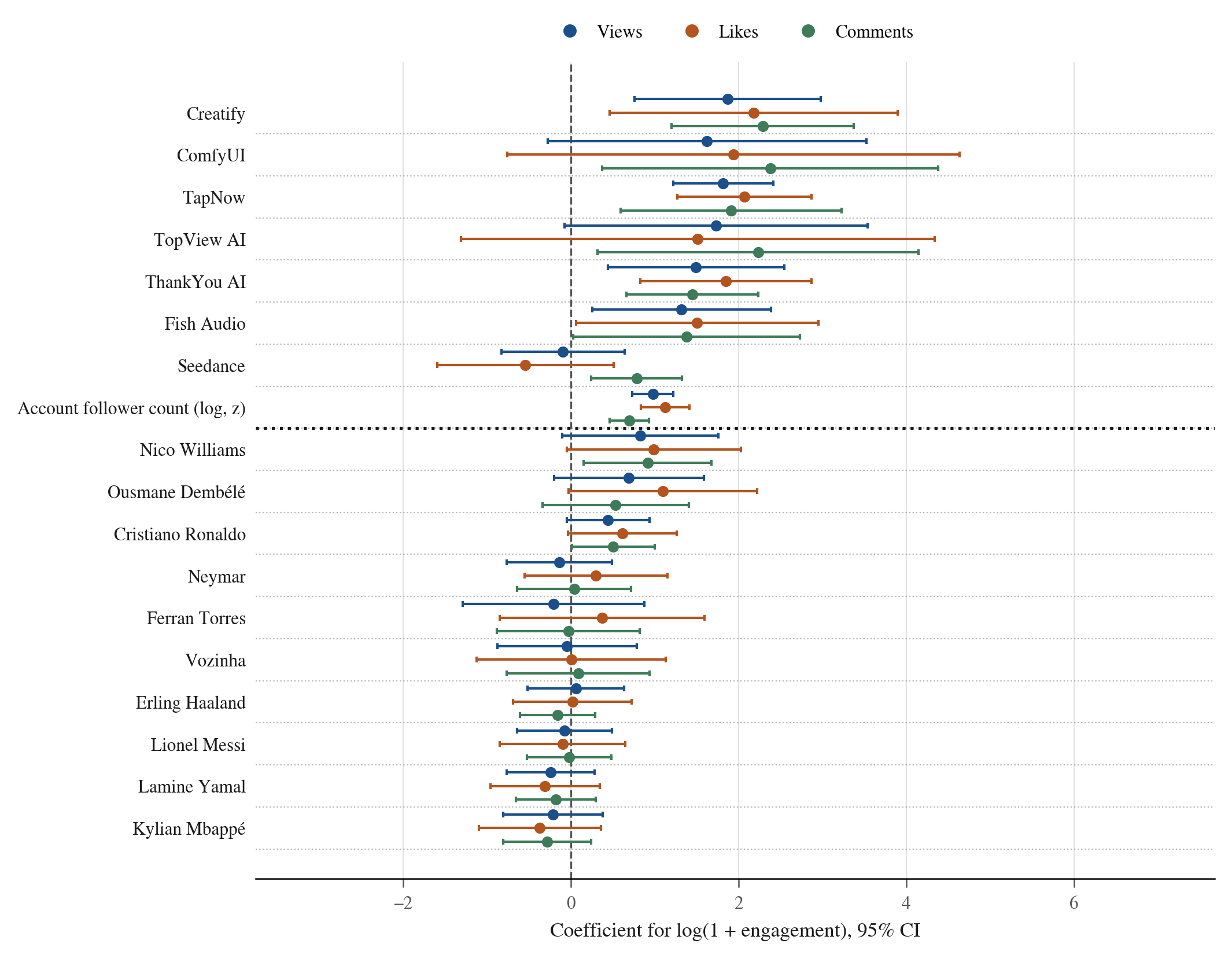}
    \caption{Associations between caption characteristics and recorded engagement. Points report regression coefficients for log-transformed views, likes, and comments, with horizontal bars indicating 95\% confidence intervals. Positive coefficients indicate higher recorded engagement conditional on the other included predictors. Estimates describe associations rather than causal effects.
}
    \label{fig:caption_engagement}
\end{figure*}

\begin{figure*}[t]
    \centering
    \includegraphics[width=0.7\linewidth]{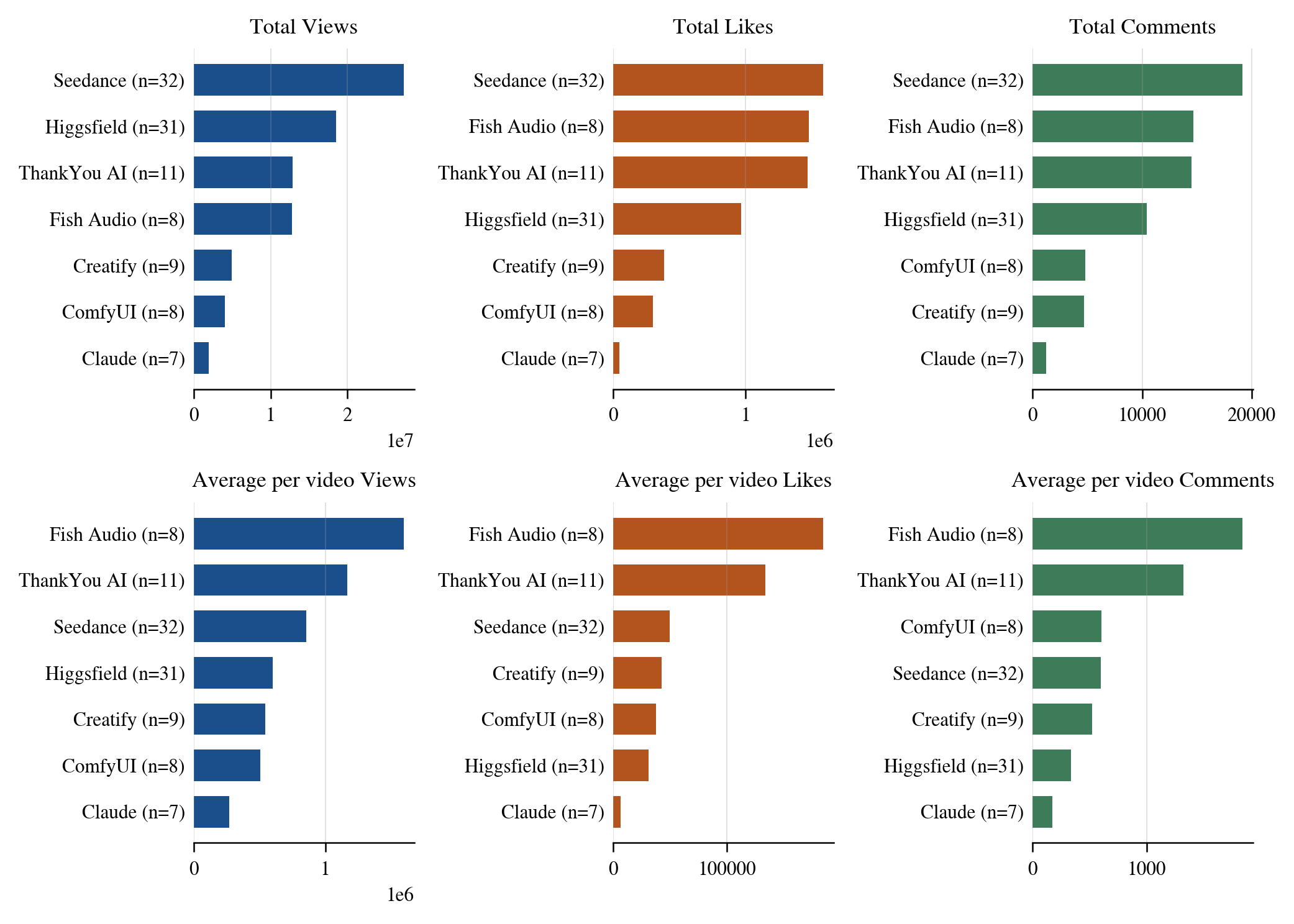}
    \caption{Aggregate and per-video engagement for posts associated with the most frequently identified generative tools. The top row reports total views, likes, and comments, while the bottom row reports average engagement per video. Tool frequencies are shown in parentheses. Differences between the two rows highlight that tools with the greatest aggregate attention are not necessarily those with the highest engagement on a per-video basis.}
    \label{fig:model_engagement_summary}
\end{figure*}

We next examined which characteristics of the posts were associated with recorded engagement. Figure~\ref{fig:caption_engagement} shows a clear distinction between generative-tool mentions and player mentions. Posts mentioning tools such as Creatify, TapNow, ThinkYou AI, and Fish Audio have substantially larger positive coefficients for views, likes, and comments than posts featuring any individual player. Creatify in particular is associated with increases across all three engagement outcomes, while several other tools show similarly consistent positive relationships. By comparison, player-level coefficients are considerably smaller and generally concentrated around zero, with confidence intervals frequently overlapping zero. Even highly prominent players such as Cristiano Ronaldo, Neymar, Lionel Messi, and Kylian Mbapp\'e do not exhibit associations comparable in magnitude to those of the leading generative tools. Within this corpus, engagement therefore appears to be more strongly associated with the model or tool attached to the content than with the football player it depicts, suggesting that the production technology itself may be an important component of audience attention.

The distinction between aggregate attention and per-video engagement is also visible across generative tools (Figure~\ref{fig:model_engagement_summary}). Seedance receives the highest total views, likes, and comments, while Higgsfield also accumulates substantial aggregate attention, reflecting in part their relatively frequent appearance in the dataset. However, the per-video rankings differ considerably. Fish Audio, which appears in only eight videos, has the highest average views, likes, and comments per video, while ThankYou AI also performs strongly on a per-video basis despite appearing in only eleven videos. In contrast, Seedance remains strong but falls behind these less frequently represented tools when engagement is normalized by the number of associated videos. This divergence shows that aggregate popularity and per-video engagement capture different aspects of model-associated attention. Because tool frequencies vary substantially across the corpus, these comparisons characterize engagement associated with tool mentions rather than the intrinsic performance of the underlying generative systems.

\begin{figure*}[t]
    \centering
    \includegraphics[width=0.7\linewidth]{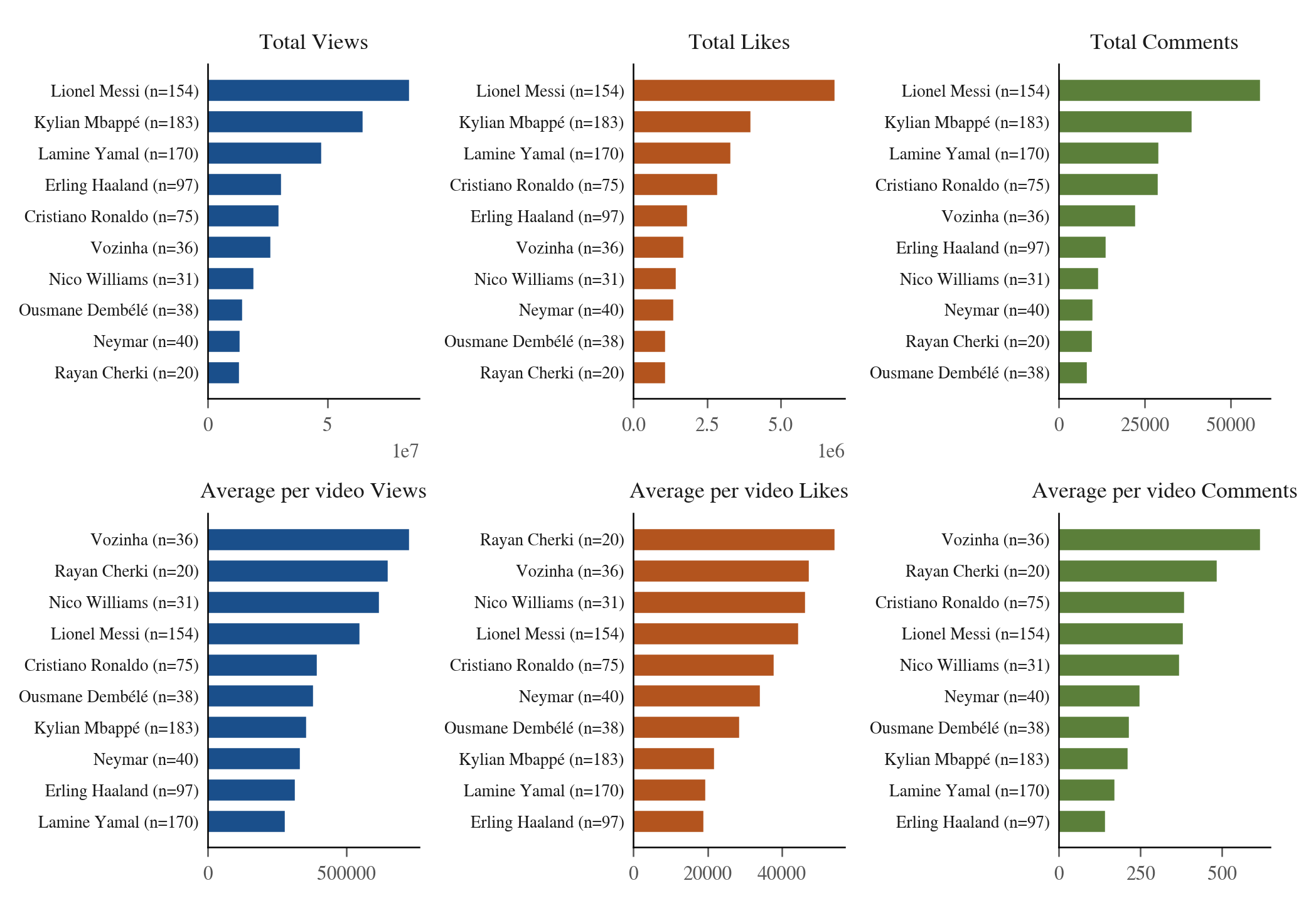}
    \caption{Aggregate and per-video engagement for posts containing the most frequently identified football players. The top row reports total views, likes, and comments, while the bottom row reports average engagement per video. The number of videos containing each player is shown in parentheses. Differences between the two rows distinguish overall accumulated attention from engagement conditional on a
player's appearance.}
    \label{fig:player_engagement_summary}
\end{figure*}

The distinction between aggregate attention and per-video engagement is also visible for player identities (Figure~\ref{fig:player_engagement_summary}). Players appearing in more videos can accumulate high total engagement without producing the highest engagement on a per-video basis. Lionel Messi, for example, leads in total views, likes, and comments, while Kylian Mbapp\'e and Lamine Yamal also rank highly in aggregate attention, reflecting in part their frequent appearance in the dataset. However, the per-video rankings differ substantially: Vozinha and Rayan Cherki, despite appearing in far fewer videos, rank among the highest in average views, likes, and comments. This divergence suggests that prominence in the corpus and engagement conditional on appearing in a video capture distinct dimensions of player-associated attention. Because the number of videos differs considerably across players, aggregate totals should therefore be interpreted alongside per-video averages rather than as direct evidence that a particular player independently drives engagement.

More broadly, the results distinguish who or what appears in the content from the attention that the resulting posts receive. Engagement is associated with a combination of sporting identities, cultural framing, posting characteristics, and account context rather than simply the prevalence of a particular athlete or anime franchise. This answers \textbf{RQ2}.

\subsection{Circulation: Sharing and Reuse}

To identify repeated content within the dataset, we developed a content-reuse detection procedure using face crops already extracted by the face-detection pipeline. The analysis used 19,094 temporally distributed face crops across 1,548 videos. Each crop was converted to a 64-bit perceptual hash (pHash), and for each video pair we calculated bidirectional best-match Hamming distances and the proportion of crops with matches at different distance thresholds. We classified two videos as belonging to the same repost group when their minimum bidirectional coverage at Hamming distance four was greater than zero.

This procedure produced 4,071 candidate video-pair links and 124 connected repost groups involving 772 videos. Manual validation of 30 random cases found no classification disagreements. The method also distinguished reused content from videos featuring the same football player but different underlying footage, supporting its use as a measure of content reuse rather than player identity. Reused material therefore represents a substantial portion of the evaluated corpus rather than a small set of isolated duplicates.

\begin{figure*}[!htb]
    \centering
    \includegraphics[width=0.7\linewidth]{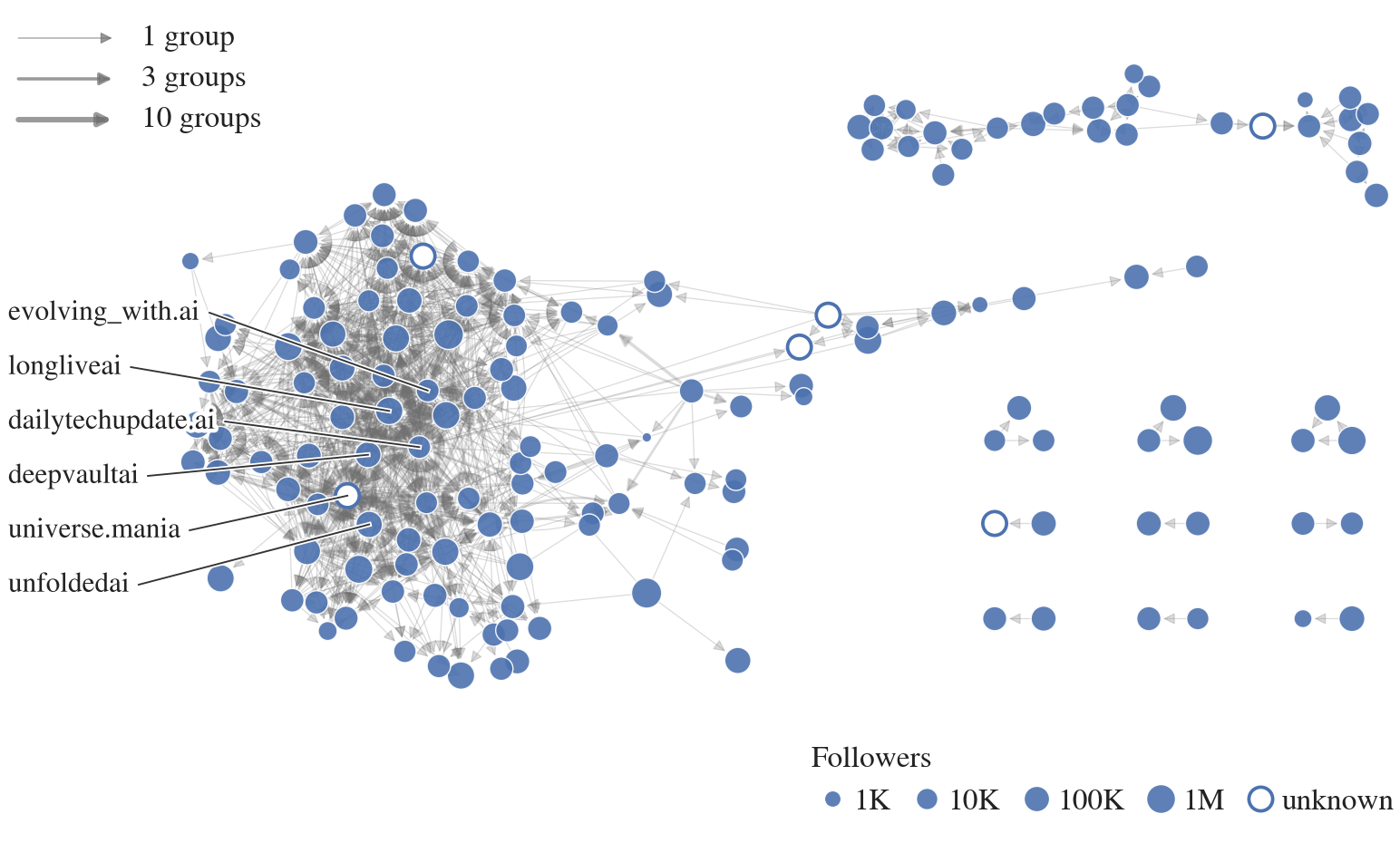}
    \caption{Account-level network of shared visual material. Nodes represent Instagram accounts, and edges connect accounts that posted videos belonging to the same detected repost group. Directed edges point from the earlier-posting account to the later-posting account within each group. Edge weights count the number of distinct repost groups supporting the relationship. The network represents temporal co-sharing rather than verified transmission or original authorship.}
    \label{fig:repost_network}
\end{figure*}

To characterize cross-account circulation, we constructed an account-level network from the detected repost groups. Of the 124 groups, 71 (57.3\%) spanned multiple known accounts, involving 167 distinct accounts. Figure~\ref{fig:repost_network} shows a large, densely interconnected component alongside a smaller component and several disconnected pairs and triads. The dense component indicates overlapping content reuse across many accounts, while the smaller components capture sharing among more limited sets of accounts. Sparsely connected accounts also extend outward from the dense component.

For each repost group, edges point from the earlier-posting account to the later-posting account, using each account's earliest observed post within that group. Edge weights count the number of distinct repost groups supporting each direction. These links indicate temporal ordering within our corpus rather than direct copying or original authorship. Together, the network shows that reuse occurs both within a broad interconnected ecosystem of accounts and within smaller, more localized groups.

\begin{figure}
    \centering
    \includegraphics[width=\linewidth]{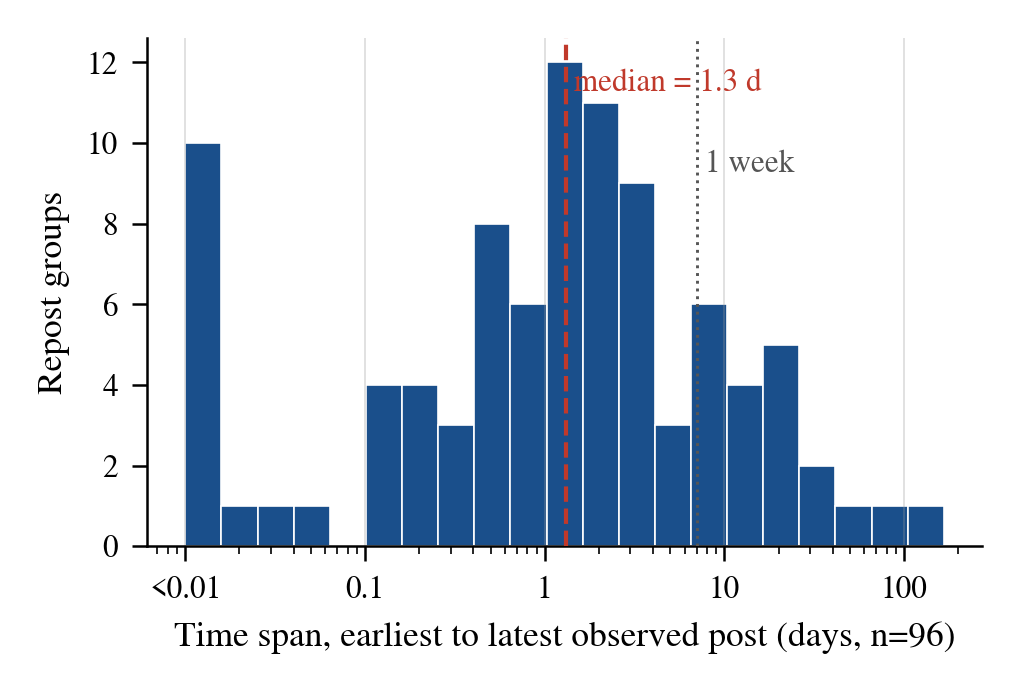}
    \caption{Observed timespans of detected content reuse. Each observation reports the interval between the earliest and latest known appearance within a repost group. The distribution is concentrated at short intervals but includes a long tail of material resurfacing over longer periods.}
    \label{fig:timing}
\end{figure}

Reposting also frequently occurred over short time intervals. Among the 96 repost groups with sufficient timestamp information, the median span between the earliest and latest observed post was 1.3 days (Figure~\ref{fig:timing}). The histogram shows that many repost groups were concentrated near the left side of the distribution, with 40 groups completing their observed repost cycle within a single day. This suggests that reused videos often circulated soon after their first observed appearance. At the same time, the log-scaled x-axis reveals a longer tail: 20 groups continued to receive reposts more than one week later, and the longest observed span was 166.3 days. Reuse therefore combines rapid circulation with longer-term resurfacing.

\begin{figure}
    \centering
    \includegraphics[width=1\linewidth]{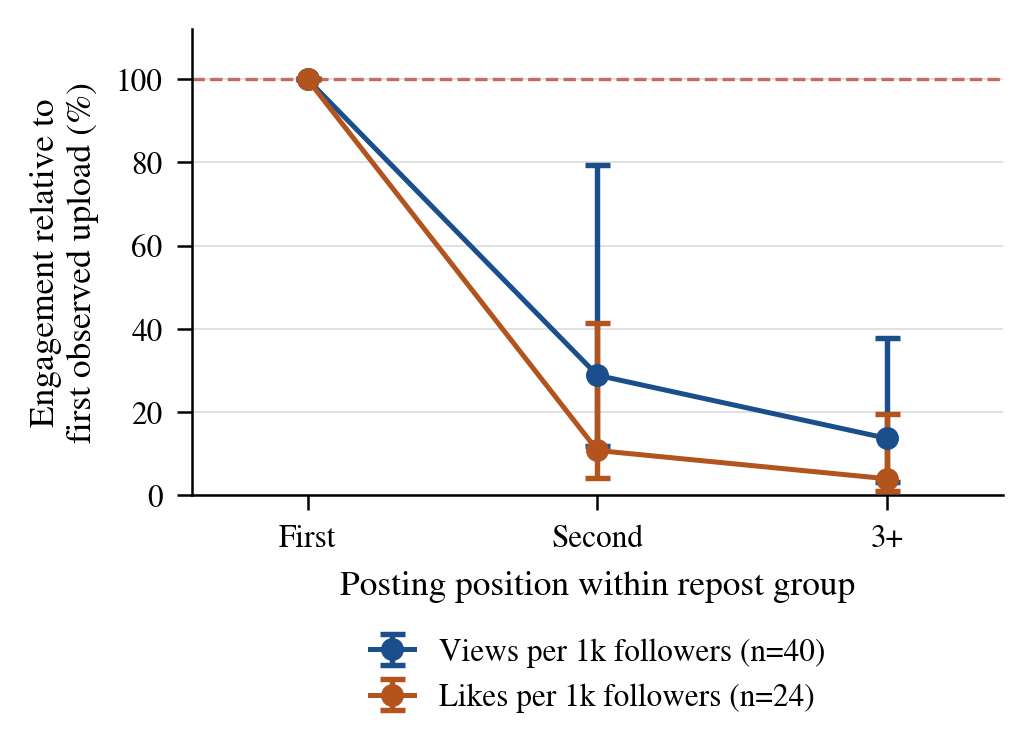}
    \caption{Engagement across successive appearances of detected reused content. Views and likes are normalized by time since posting and follower count and expressed relative to the first observed post within each repost group. ``First'' denotes the earliest appearance observed in the corpus rather than verified original authorship.}
    \label{fig:repost_relative_engagement}
\end{figure}

We next examined whether engagement declined as the same content was reposted across accounts. Because repost groups often involved accounts with substantially different audience sizes, we normalized engagement by both time since posting and follower count, then expressed each post relative to the first observed post in its repost group.

As shown in Figure~\ref{fig:repost_relative_engagement}, engagement declined sharply after the first observed post. The second post received a median of 29\% of the first post's views per day per follower and 11\% of its likes per day per follower. For posts appearing third or later, these values declined further to 14\% and 4\%, respectively. Although uncertainty was larger for second posts, particularly for likes, the overall pattern remained consistent: later reposts generated substantially lower engagement relative to audience size than the first observed appearance of the same content.

These patterns do not establish a causal penalty for reposting. Later appearances can differ from earlier ones in account characteristics, posting context, and opportunities to accumulate engagement, and the earliest observed post need not be the original source. Nevertheless, the results demonstrate that the circulation of identical or closely related content is associated with strongly unequal attention across successive appearances.

Taken together, the reuse analysis shows that nearly half of evaluated videos belong to detected reuse groups, that shared material connects a broad network of accounts, that circulation occurs on both short and long timescales, and that later observed appearances receive substantially less normalized engagement. Cultural visibility in this corpus therefore depends not only on what creators produce, but also on which material is redistributed and where in its observed circulation sequence a post appears. This answers \textbf{RQ3}.

\section{Discussion}

Generative AI expands the supply of cultural production, but our results show that abundance does not necessarily translate into representational diversity. A relatively small set of football figures dominate visual representation, while \emph{Blue Lock}, \emph{Naruto}, and \emph{One Piece} account for 60.1\% of franchise mentions. This parallels experimental evidence that generative AI can increase individual creative output while reducing diversity across outputs \citep{doshi2024creativity}. Our data cannot establish that AI caused this concentration, but they show that a diverse technical ecosystem can coexist with a narrow cultural repertoire. In this sense, lower production costs may expand opportunities to create without fundamentally redistributing which people and cultural references receive visibility.

The football--anime crossover also illustrates a function of synthetic media that extends beyond deception. Athletes are not simply rendered in a different visual style; they become material for fictional representation. This builds on real-person fiction, where public figures and their mediated personas become characters for participatory storytelling \citep{waysdorf2015football,popova2017persona}. Generative media make this form of participation increasingly visual and potentially responsive to unfolding events. The normative question is therefore not only whether synthetic content is mistaken for reality, but who can construct and circulate public portrayals of recognizable people. Our current measures identify the athletes and fictional resources involved, but direct measurement of event-to-character or hero--villain mappings remains an important next step.

Supply and attention are also distinct. The most common references in the corpus are not uniformly associated with greater engagement, while contextual and caption-derived features explain more variation in engagement than the selected visual identities alone. This extends evidence that AI provenance itself provides an incomplete account of audience response: cultural form and human curation shape how synthetic media perform \citep{chang2026meme,weener2026bluebird}. Generative content enters existing fandoms, events, and platform conventions rather than competing on technical novelty alone. More broadly, this reinforces the distinction between expanded access to cultural production and access to attention, which remains uneven and platform-mediated \citep{nieborg2018platformization}.

Finally, creation is only part of how this material becomes visible. Approximately half of evaluated videos belong to detected reuse groups, many spanning multiple accounts, and later observed appearances receive substantially less normalized engagement. Repetition and redistribution are longstanding components of participatory culture and meme circulation \citep{jenkins2013spreadable,shifman2013memes,zulli2022imitation}; our results show that they are also central to generative video. The same apparent prominence can therefore arise from many distinct creations or from repeated circulation of relatively little material. Taken together, the findings suggest that the social consequences of generative media emerge across three linked but unequal stages: what people can produce, what audiences attend to, and what networks continue to circulate.

\bibliography{references}

\appendix
\section{Appendix}
\label{sec:supplementary}

\subsection{Visual Identity Diagnostics}

\begin{figure}[!t]
    \centering
    \includegraphics[width=\columnwidth]
    {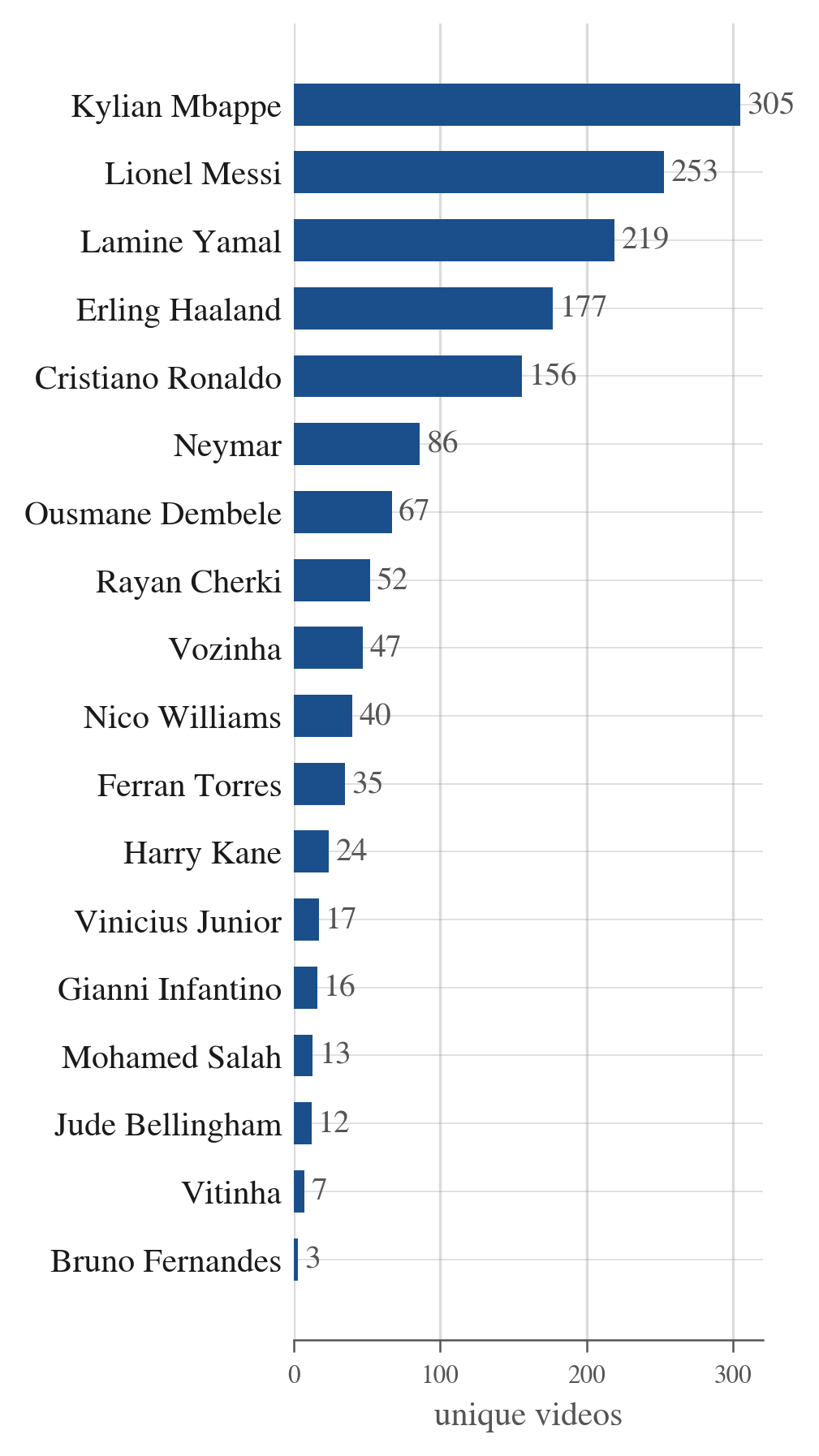}
    \caption{Complete player-level coverage in the visual analysis.
    Bars report the number of unique videos containing each identified
    player. Unlike Figure~2, all 18 identified figures are shown
    separately.}
    \label{fig:appendix-player-coverage}
\end{figure}

Figure~\ref{fig:appendix-player-coverage} provides the complete
player-level distribution underlying the aggregated results reported
in Figure~2. Among the 18 identified figures, Kylian Mbappé appeared
in the largest number of unique videos, followed by Lionel Messi,
Lamine Yamal, Erling Haaland, and Cristiano Ronaldo. Representation
declined substantially beyond these frequently depicted players.

\begin{figure}[!t]
    \centering
    \includegraphics[width=\columnwidth]
    {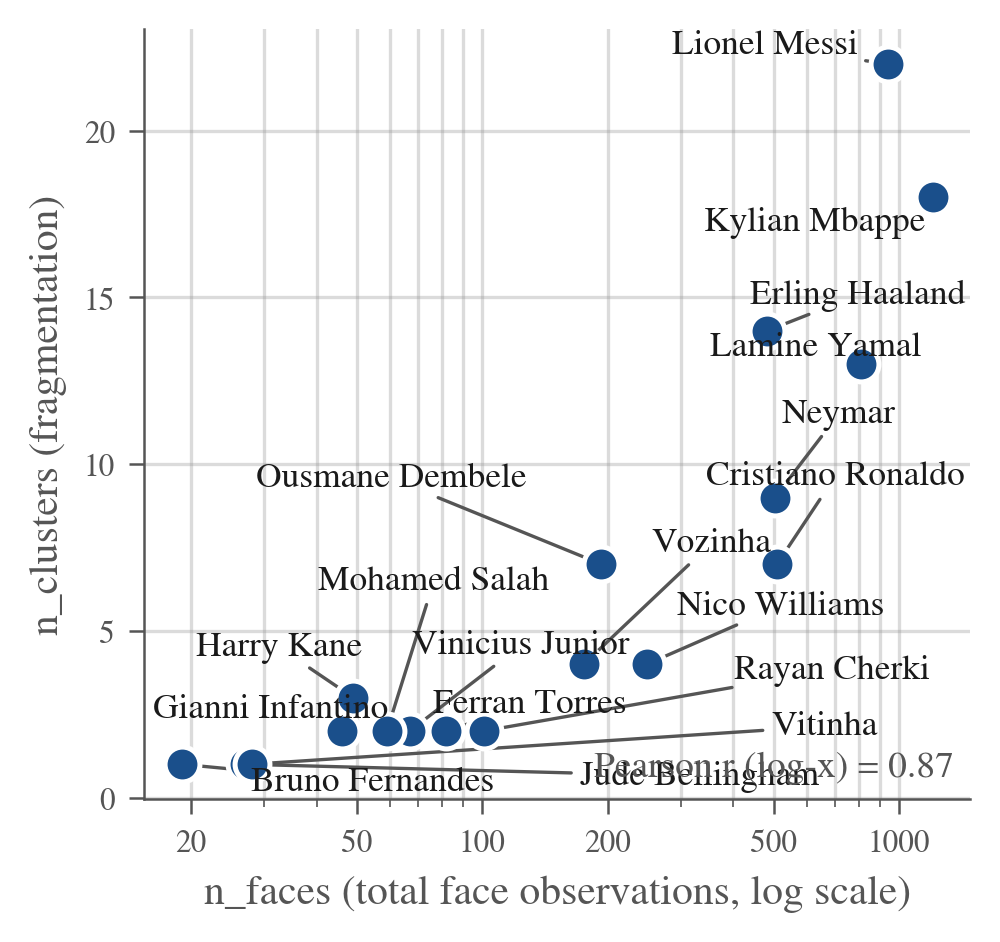}
    \caption{Relationship between the number of retained face
    observations and the number of recovered clusters assigned to
    each identified player. The horizontal axis is logarithmic.
    Players with more observations generally appear across more
    clusters ($r=0.87$).}
    \label{fig:cluster-fragmentation}
\end{figure}

We also examined the relationship between the number of face
observations associated with a player and the number of recovered
clusters assigned to that player. As shown in
Figure~\ref{fig:cluster-fragmentation}, players with more face
observations generally have more recovered clusters, with a strong
positive association ($r=0.87$). The horizontal axis is displayed
on a logarithmic scale to accommodate the substantial differences
in observation volume across players.

The relationship is not proportional across players. Lionel Messi
and Kylian Mbappé have the largest numbers of face observations and
also the greatest numbers of clusters, while several players with
substantially fewer observations are represented by only one or two
clusters. The increasing number of clusters with observation volume
indicates that part of the apparent fragmentation is attributable to
the amount of visual material available for a player. Raw cluster
counts should therefore be interpreted relative to observation
volume rather than treated alone as evidence of greater visual
variation.

\begin{figure}[!t]
    \centering
    \includegraphics[width=\columnwidth]
    {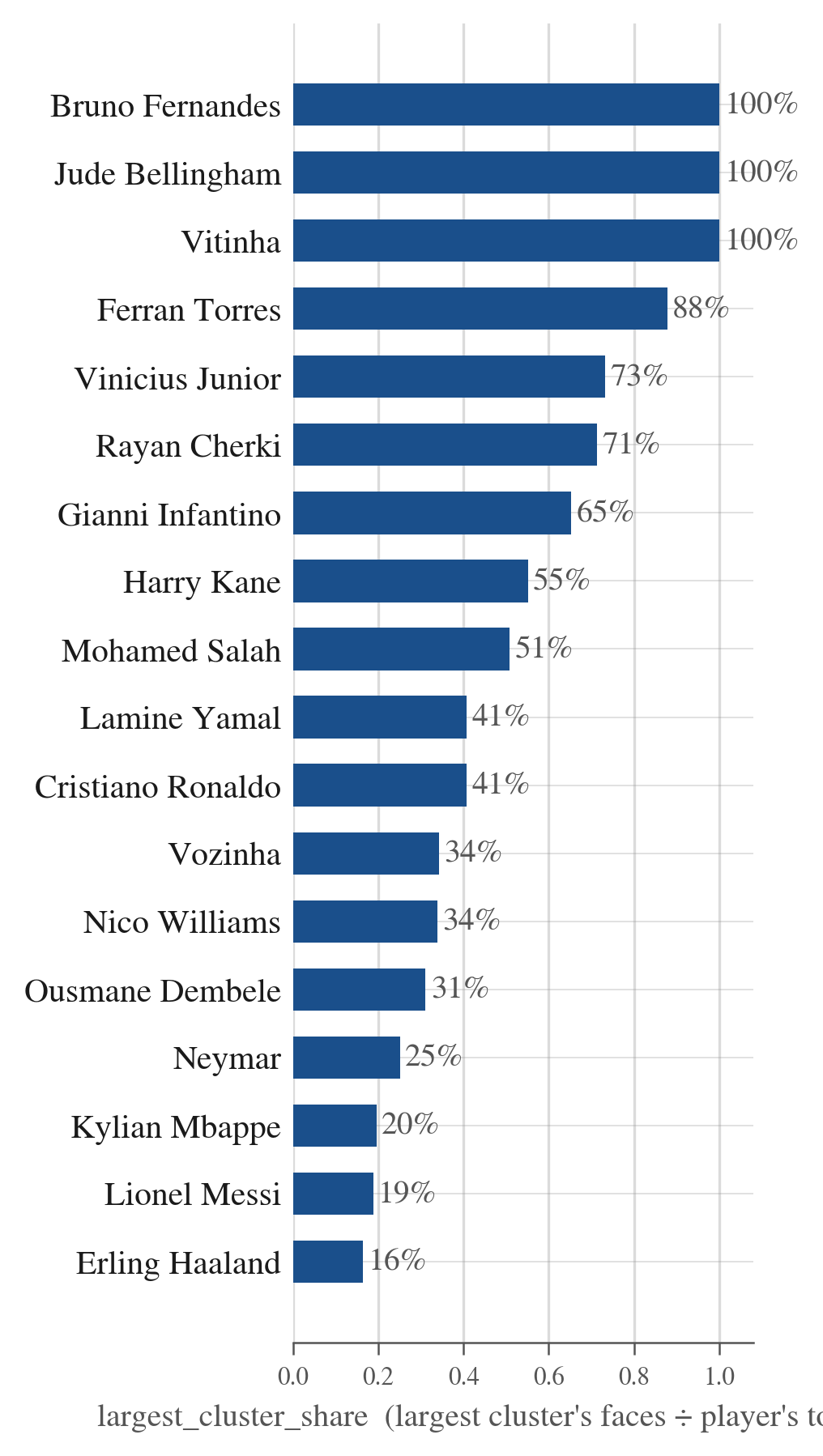}
    \caption{Share of each player's labeled face observations
    contained in that player's largest recovered cluster. Higher
    values indicate greater concentration in a single cluster,
    whereas lower values indicate that observations were divided
    across several clusters.}
    \label{fig:largest-cluster-share}
\end{figure}

Figure~\ref{fig:largest-cluster-share} provides a complementary
measure of fragmentation by reporting the proportion of each
player's observations contained in that player's largest cluster.
Lower values indicate that observations were distributed across
several clusters, while higher values indicate that one cluster
accounted for most of the player's observations. Players with few
observations can have high shares simply because they were assigned
to only one or two clusters. This measure should therefore be
interpreted alongside total observation volume.

\FloatBarrier

\subsection{Caption-Derived Supply Results}

\subsubsection{Named Generative Tools}

\begin{figure}[!t]
    \centering
    \includegraphics[width=\columnwidth]
    {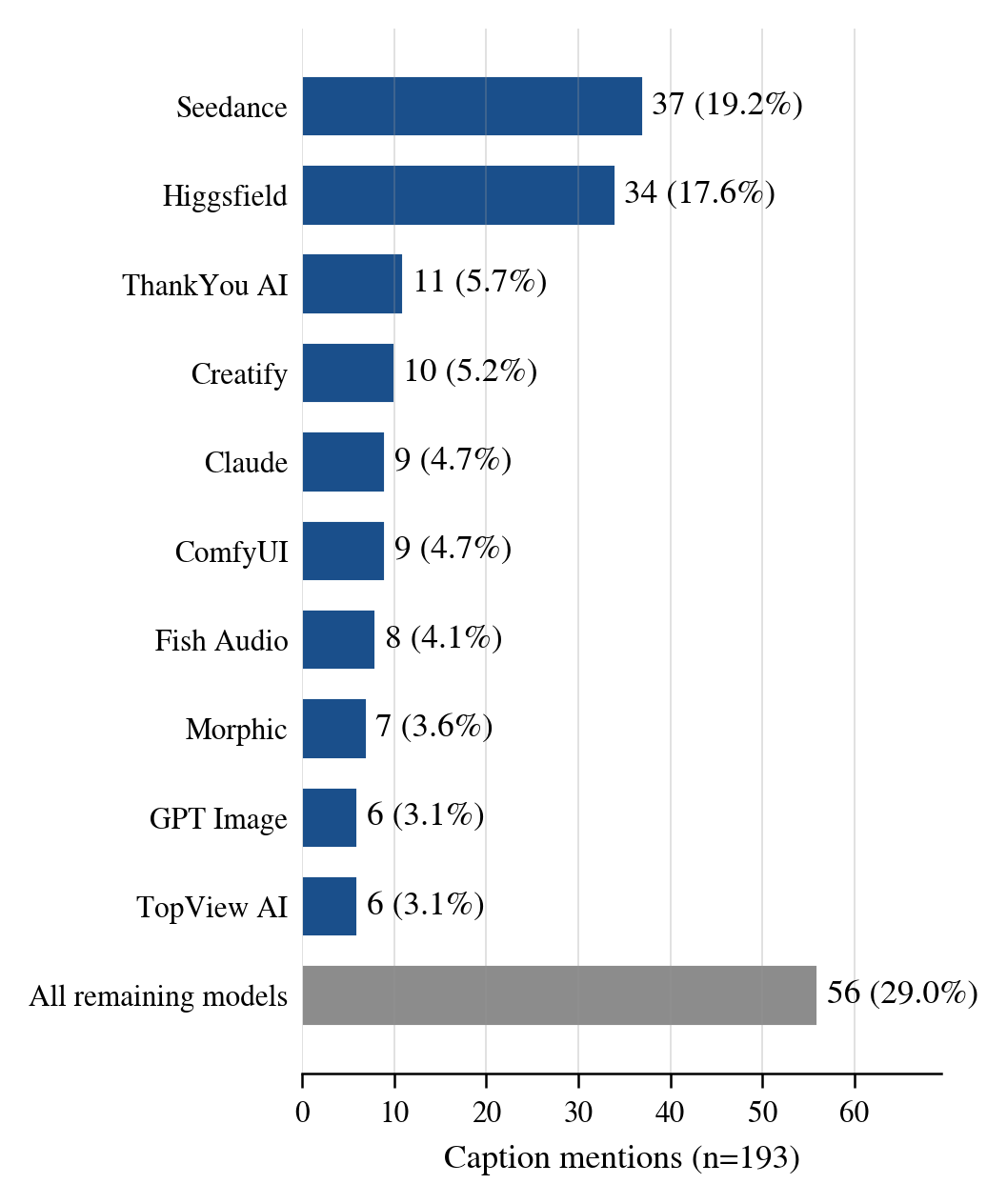}
    \caption{Explicit generative-tool mentions extracted from
    captions and hashtags ($n=193$ mentions). Tools outside the ten
    most frequently named are aggregated into ``All remaining
    models.''}
    \label{fig:model-mentions}
\end{figure}

Captions and hashtags contained 193 explicit generative-tool
mentions. Seedance was the most frequently named tool, with 37
mentions (19.2\%), followed by Higgsfield with 34 (17.6\%).
ThankYou AI appeared 11 times (5.7\%) and Creatify appeared 10
times (5.2\%), while Claude and ComfyUI were each named nine times
(4.7\%). Fish Audio, Morphic, GPT Image, and TopView AI each
accounted for fewer than 5\% of mentions. The remaining 56 mentions
(29.0\%) were distributed across less frequently named tools
(Figure~\ref{fig:model-mentions}).

These counts measure explicit textual references. They do not
independently verify which system was used to produce a video.

\subsubsection{Attributed Origins of Named Tools}

\begin{figure}[!t]
    \centering
    \includegraphics[width=\columnwidth]
    {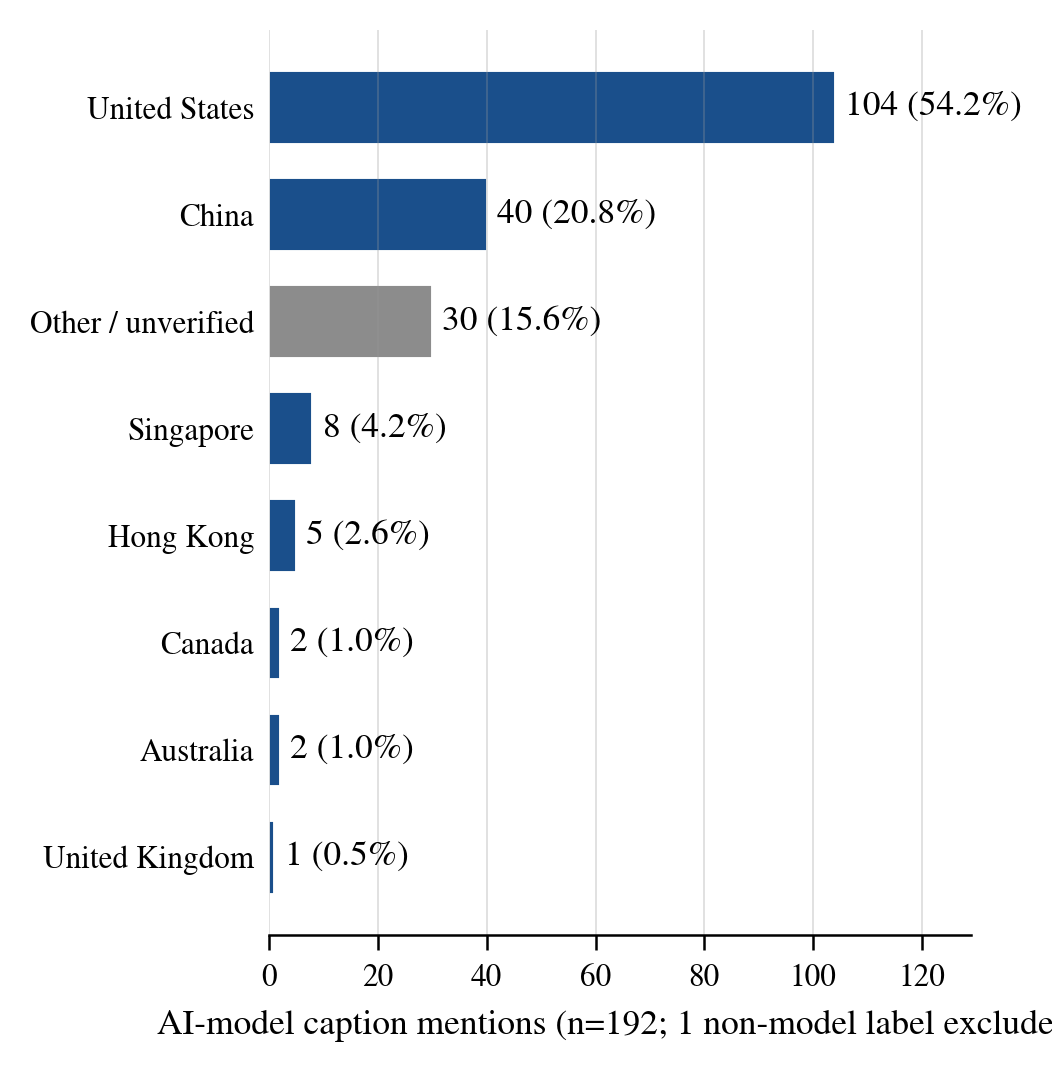}
    \caption{Attributed geographic origins of generative tools named
    in captions and hashtags ($n=192$ coded mentions). One extracted
    non-tool label was excluded. Country assignments refer to the
    named tools rather than the posting accounts or content creators.}
    \label{fig:model-origin}
\end{figure}

After excluding one extracted label that did not correspond to a
generative tool, 192 tool mentions were assigned to geographic-origin
categories. Tools associated with the United States accounted for
104 mentions (54.2\%), followed by China with 40 (20.8\%).
A further 30 mentions (15.6\%) could not be verified or were grouped
into the other category. The remaining mentions were associated with
Singapore, Hong Kong, Canada, Australia, and the United Kingdom
(Figure~\ref{fig:model-origin}). These categories describe the
attributed origins of the named tools, not the locations of the
accounts or creators posting the videos.

\FloatBarrier

\subsubsection{Anime-Franchise Mentions}

\begin{figure}[!t]
    \centering
    \includegraphics[width=\columnwidth]
    {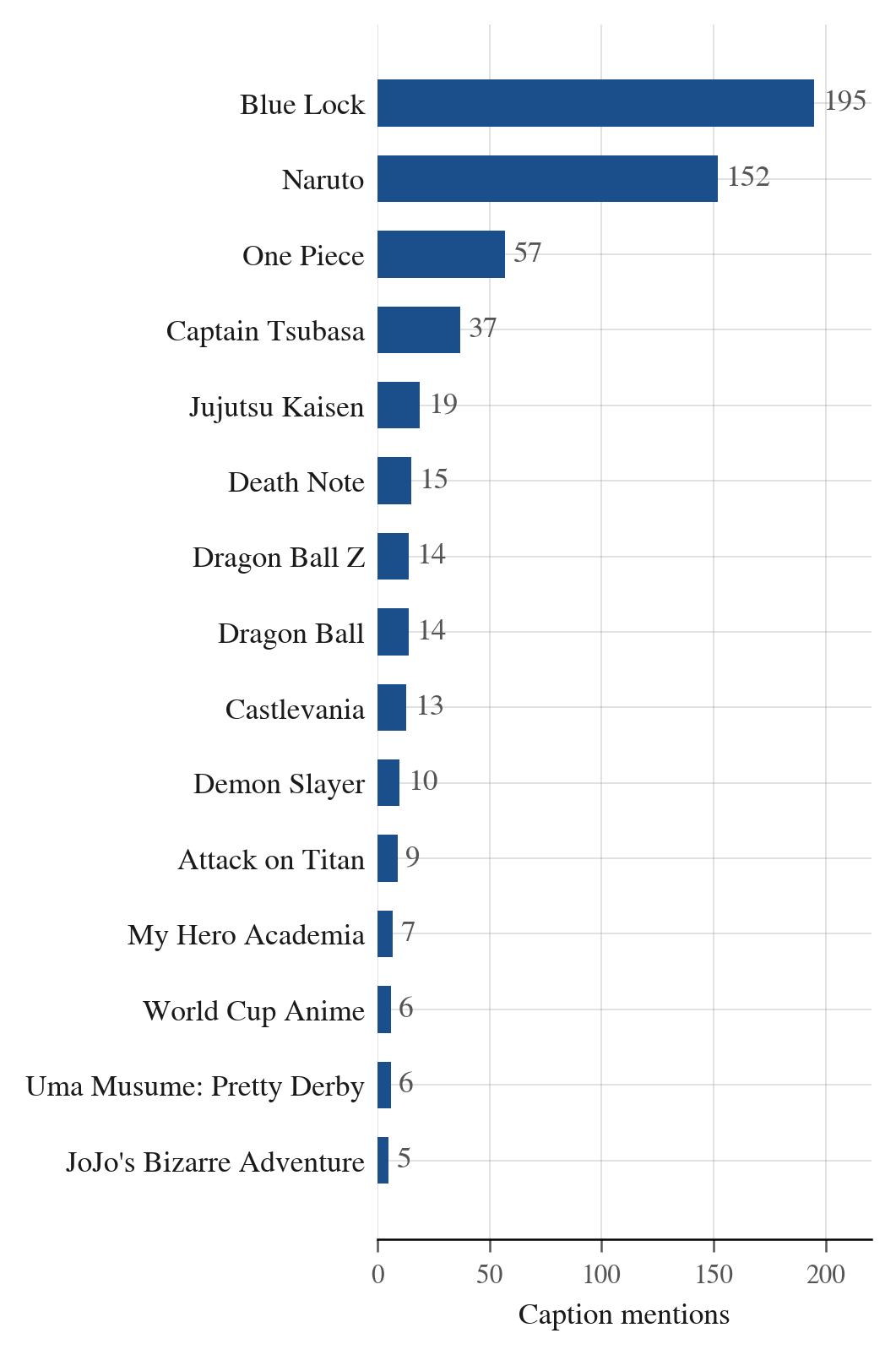}
    \caption{The 15 most frequently mentioned anime franchises in
    captions and hashtags. Counts represent explicit textual mentions;
    less frequently mentioned titles are not displayed.}
    \label{fig:anime-title-detail}
\end{figure}

Figure~\ref{fig:anime-title-detail} expands the franchise distribution
reported in Figure~3 by separately displaying the 15 most frequently
named titles. The figure reveals the additional titles contained
within the aggregated ``All other titles'' category while preserving
the strong concentration around \textit{Blue Lock},
\textit{Naruto}, and \textit{One Piece}. These counts represent
explicit caption or hashtag mentions rather than visual verification
of a video's artistic style.

We additionally combined the visually identified player labels with
the caption-derived franchise mentions. A video contributes to a cell
in Figure~\ref{fig:player_franchise_overlap} when the player was
identified in its visual content and the corresponding franchise was
explicitly named in its caption or hashtags. The heatmap therefore
measures cross-modal co-occurrence rather than establishing that the
player was depicted in the visual style or fictional setting of that
franchise.

\begin{tcolorbox}[
    title={Caption Entity Extraction Prompt},
    colback=gray!5,
    colframe=black!60,
    fonttitle=\bfseries,
    fontupper=\small,
    breakable,
    boxrule=0.5pt,
    arc=2mm,
    left=2mm,
    right=2mm
]
\textbf{System prompt}

\medskip
Normalization rules:
\begin{itemize}
    \item Normalize obvious aliases and spelling variants to a
    canonical common name, e.g.\ ``Messi'' $\rightarrow$
    ``Lionel Messi'', ``PSG'' $\rightarrow$
    ``Paris Saint-Germain'', and ``CR7'' $\rightarrow$
    ``Cristiano Ronaldo''.
    \item Keep players separate from managers and pundits.
    \item Keep national teams separate from club teams.
    \item Keep anime characters separate from real football players.
    \item Use \texttt{[]} for any field with nothing found.
    Never omit a field or use \texttt{null}.
\end{itemize}

Return ONLY a single JSON object with exactly these 13 keys,
with no other text or Markdown fences:

\begin{verbatim}
{
  "model_names": [],
  "model_versions": [],
  "player_names": [],
  "country_names": [],
  "anime_titles": [],
  "anime_characters": [],
  "studio_names": [],
  "club_team_names": [],
  "match_events": [],
  "tournament_rounds": [],
  "source_handles": [],
  "manager_pundit_names": [],
  "venue_names": []
}
\end{verbatim}

\tcblower
\textbf{User prompt template}

\medskip
Caption text:\\
\texttt{\{caption\_text\}}

\medskip
Hashtags:\\
\texttt{\{hashtag\_str\}}

\medskip
Extract the 13 fields as instructed.
Return ONLY the JSON object.

\medskip
\textit{Template note:}
\texttt{\{hashtag\_str\}} is the post's hashtags joined
as \texttt{\#tag1, \#tag2}, or \texttt{(none)} if the post
has no hashtags.
\end{tcolorbox}

\end{document}